\documentclass[10pt,conference]{IEEEtran}\IEEEoverridecommandlockouts
\usepackage{epsfig,graphicx,subfigure,psfrag,amsmath,cases}
\usepackage{latexsym,amssymb,amsmath,epsfig,subfigure,algorithm,mathtools}
\usepackage{algorithmic}
\usepackage{color}
\usepackage{url}
\usepackage{scrtime}
\usepackage{array}

\author{\authorblockN{Derrick Wing Kwan Ng and Robert Schober\thanks{The authors are also with the University of British Columbia. This work was supported in part by the AvH Professorship Program of the Alexander von Humboldt Foundation.}}
Institute for Digital Communications\\
Friedrich-Alexander-University Erlangen-N\"urnberg (FAU), Germany}

\title{Resource Allocation for Coordinated Multipoint Networks with Wireless Information and Power Transfer}

\date{\thistime,\,\today}

\newtheorem{Thm}{Theorem}

\newtheorem{Remark}{Remark}
 \newcommand{\qed}{\hfill \ensuremath{\blacksquare}}
 \DeclarePairedDelimiter{\ceil}{\lceil}{\rceil}
\DeclareMathOperator{\Tr}{\mathrm{Tr}}
\DeclareMathOperator{\Rank}{\mathrm{Rank}}

\DeclareMathOperator{\diag}{\mathrm{diag}}
\DeclareMathOperator{\vect}{\mathrm{vec}}
\DeclareMathOperator{\maxo}{maximize}
\DeclareMathOperator{\mino}{minimize}
\DeclareMathOperator{\bigo}{\cal O}
\newcommand{\abs}[1]{\lvert#1\rvert}
\newcommand{\norm}[1]{\lVert#1\rVert}
\newcommand{\bnorm}[1]{\Big\lVert#1\Big\rVert}
\newcolumntype{L}{>{\arraybackslash\raggedright}m{6cm}}

\begin{document}

\maketitle

\begin{abstract}
This paper studies the resource allocation algorithm design  for multiuser coordinated multipoint (CoMP) networks  with simultaneous wireless information and power transfer (SWIPT).  In particular, remote radio heads (RRHs) are connected to a central processor (CP) via capacity-limited backhaul links to facilitate CoMP joint transmission. Besides, the CP transfers energy to the RRHs
for more efficient network operation.  The considered resource allocation algorithm design is formulated as a non-convex optimization problem with  a minimum required signal-to-interference-plus-noise ratio (SINR) constraint at multiple  information receivers and a minimum required power transfer constraint at the energy harvesting receivers. By optimizing the transmit beamforming vectors at the CP and energy sharing between the CP and the RRHs, we aim at jointly minimizing the total network transmit power and the maximum capacity consumption per backhaul  link.
The resulting non-convex optimization problem is NP-hard. In light of the intractability of the problem, we reformulate it by replacing the non-convex objective function with its convex hull, which enables the derivation of an efficient
iterative resource allocation algorithm. In each iteration, a non-convex optimization problem is solved by  semi-definite programming (SDP) relaxation and the proposed iterative algorithm converges to a local optimal solution of the original problem. Simulation results illustrate that our proposed algorithm achieves a close-to-optimal performance and provides a significant reduction in  backhaul capacity consumption compared to full cooperation. Besides, the considered CoMP network is shown to provide superior system performance as far as  power consumption is concerned compared to a traditional system with  multiple antennas co-located.
\end{abstract}

\renewcommand{\baselinestretch}{0.92}
\large\normalsize

\section{Introduction} \label{sect1}
Next generation wireless communication networks are required to provide ubiquitous and high data rate communication  with guaranteed quality of service (QoS). These requirements have led to a tremendous need for energy in both transmitter(s) and receiver(s). In practice, portable mobile devices are
typically powered by capacity limited batteries which require frequent recharging. Besides,  battery technology has developed very slowly  over the past decades and the battery capacities available in the near future will be unable to improve this situation. Consequently, energy harvesting based mobile communication system design has become
a prominent approach for addressing this issue. In particular, it enables self-sustainability for energy
limited communication networks. In addition to conventional energy harvesting sources such as solar, wind, and biomass, wireless power transfer  has been proposed as an emerging alternative energy source, where the receivers scavenge energy from the ambient radio frequency (RF)  signals \cite{CN:Shannon_meets_tesla}--\nocite{JR:MIMO_WIPT,JR:WIPT_fullpaper}\cite{JR:Kwan_secure_imperfect}. In fact,  wireless power transfer technology not only eliminates the need of power cords and chargers, but also facilitates one-to-many charging due to the broadcast nature of wireless channels. More importantly, it enables the possibility  of simultaneous wireless information and power transfer (SWIPT) leading to many interesting and challenging new research problems which have to be solved  to bridge the gap between theory and practice. In \cite{CN:Shannon_meets_tesla}, the authors investigated the fundamental trade-off between  harvested energy and wireless channel capacity across a pair of coupled  inductor circuit in the presence of   additive
white Gaussian noise. Then, in \cite{JR:MIMO_WIPT},  the study was extended  to  multiple antenna wireless broadcast systems.
In   \cite{JR:WIPT_fullpaper},  the energy efficiency of  multi-carrier systems  with  SWIPT was revealed.  Specifically, it was shown in \cite{JR:WIPT_fullpaper} that integrating an energy harvester into a conventional information receiver  improves the  energy efficiency of a communication network. In \cite{JR:Kwan_secure_imperfect}, robust beamforming design for SWIPT systems with physical layer security was investigated. The results in \cite{CN:Shannon_meets_tesla}--\cite{JR:Kwan_secure_imperfect} indicate that both the information rate and the amount of harvested energy at the receivers can be significantly increased at the expense of an increase in the transmit power. However, despite the promising results in the literature, the performance of wireless power/energy transfer systems is still limited by  the distance between the transmitter and the receiver due to the  high signal attenuation associated with path loss.

Coordinated multipoint (CoMP) transmission is an important technique for extending service coverage, improving spectral efficiency, and  mitigating interference \cite{JR:comp}\nocite{JR:comp2,JR:limited_backhaul,JR:Quek}--\cite{CN:Wei_yu_sparse_BF}. A possible deployment scenario for  CoMP networks  is to split the functionalities of the base stations between a central processor (CP)  and a set of remote
radio heads (RRHs). In particular, the  CP performs the power hungry and computationally intensive baseband signal processing while the RRHs are responsible for all radio frequency (RF) operations such as analog filtering and power amplification. Besides, the RRHs are distributed across the network and connected to the CP via backhaul links.   This system architecture is known as cloud computing network. As a result, the CoMP systems architecture inherently
provides  spatial diversity   for combating  path loss and shadowing. It has been shown  that a significant system performance gain  can be achieved when full cooperation is enabled in CoMP systems  \cite{JR:comp,JR:comp2}. However, in practice, the enormous signalling overhead incurred by the information exchange between the CP and the RRHs may be infeasible when the capacity of the backhaul link is limited. Hence, resource allocation for CoMP networks  with finite backhaul capacity has attracted much
attention in the research community \cite{JR:limited_backhaul}--\cite{CN:Wei_yu_sparse_BF}. In \cite{JR:limited_backhaul}, the authors studied the energy efficiency of CoMP multi-cell networks with capacity constrained backhaul links. In \cite{JR:Quek} and \cite{CN:Wei_yu_sparse_BF}, iterative sparse beamforming algorithms were proposed to reduce the load of the backhaul links while providing reliable communication to the users. However, the energy sources of the receivers
in \cite{JR:comp}\nocite{JR:comp2,JR:limited_backhaul,JR:Quek}--\cite{CN:Wei_yu_sparse_BF} were assumed to be  perpetual and this assumption may not be valid for power-constrained portable
devices. On the
other hand, the  signals transmitted by the RRHs could be exploited for energy harvesting by
the power-constrained receivers for extending
their  lifetimes. However, the resource allocation algorithm design for  CoMP  SWIPT systems has not been solved sofar, and will be tackled in this paper.

Motivated by the aforementioned observations, we formulate the resource allocation algorithm design for multiuser CoMP communication networks  with SWIPT as a non-convex optimization problem. We  jointly minimize  the total network transmit power and the maximum capacity consumption per backhaul link while ensuring quality of service (QoS) for reliable communication and efficient wireless power transfer.  In particular, we propose an iterative algorithm which provides a local optimal solution for the considered optimization problem.

\section{System Model}
\label{sect:system model}

\subsection{Notation}
We use boldface capital and lower case letters to denote matrices and vectors, respectively. $\mathbf{A}^H$, $\Tr(\mathbf{A})$, and $\Rank(\mathbf{A})$ represent the  Hermitian transpose, trace, and rank of  matrix $\mathbf{A}$; $\mathbf{A}\succ \mathbf{0}$ and $\mathbf{A}\succeq \mathbf{0}$ indicate that $\mathbf{A}$ is a positive definite and a  positive semidefinite matrix, respectively; $\vect(\mathbf{A})$ denotes the vectorization of matrix $\mathbf{A}$ by stacking its columns from left to right to form a column vector; $\mathbf{I}_N$ is the $N\times N$ identity matrix; $\mathbb{C}^{N\times M}$ denotes the set of all $N\times M$ matrices with complex entries; $\mathbb{H}^N$ denotes the set of all $N\times N$ Hermitian matrices; $\diag(x_1, \cdots, x_K)$ denotes a diagonal matrix with the diagonal elements given by $\{x_1, \cdots, x_K\}$; $\abs{\cdot}$ and $\norm{\cdot}_p$ denote the absolute value of a complex scalar and the
$l_p$-norm of a vector, respectively. In particular, $\norm{\cdot}_0$ is known as the $l_0$-norm of a vector and denotes the number of non-zero entries in the vector; the circularly symmetric complex Gaussian (CSCG) distribution is denoted by ${\cal CN}(\mu,\sigma^2)$ with mean  $\mu$ and variance $\sigma^2$; $\sim$ stands for ``distributed as"; $\ceil[\big]{x}$ is the ceiling function denoting the smallest integer not smaller than $x$.

\subsection{CoMP Network Model and Central Processor}
\label{sect:multicell-central-unit}
\begin{figure}
\centering
\includegraphics[width=3.5in]{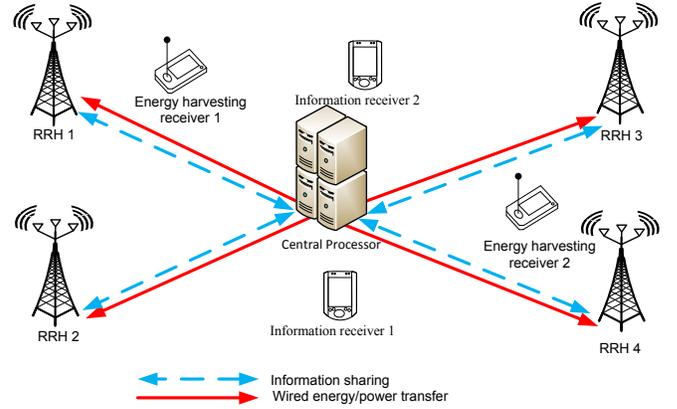}\vspace*{-2mm}
\caption{Coordinated multipoint (CoMP) multiuser downlink communication system model with a central processor (CP), $L=4$ remote radio heads (RRHs), $K=2$  information receivers (IRs), and $M=2$ energy harvesting receiver (ERs).}
\label{fig:system_model}\vspace*{-4mm}
\end{figure}

We consider a CoMP multiuser downlink communication network. The system consists of a CP,  $L$ RRHs,   $K$ information receivers (IRs), and $M$ energy harvesting receivers (ERs), cf. Figure \ref{fig:system_model}. Each RRH is equipped with $N_\mathrm{T}>1$ transmit antennas. The IRs and ERs are single antenna devices  which exploit the received signal powers in the RF for information decoding and energy harvesting, respectively. In practice, the ERs may be idle IRs which are scavenging  energy from the RF for extending their lifetimes. On the other hand, the CP is the core unit in the network. In particular, it has the data of all information receivers. Besides, we assume that the global channel state information (CSI) is perfectly known at
the CP and all computations are performed in this
unit. Based on the available CSI, the CP computes the
resource allocation policy and broadcasts it to  all RRHs. Specifically, each RRH receives the control signals for resource allocation and the data of the $K$ IRs from the CP via a backhaul\footnote{In practice, the backhaul links can be implemented by different technologies such as digital subscriber line (DSL) or out-of-band microwave links. } link. Furthermore, we assume that the  CP  supplies energy to  the RRHs in the network  via  dedicated power lines to support the RRHs' power consumption.


\subsection{Channel Model}

We focus on a frequency flat fading channel and a time division duplexing (TDD) system. Each RRH obtains the local CSI of all receivers by exploiting channel reciprocity and handshaking signals. Subsequently, the RRHs feed their local CSI to the CP for computation of the resource allocation policy. The received signals at IR $k\in\{1,\ldots,K\}$ and ER $m\in\{1,\ldots,M\}$ are given by
\begin{eqnarray}
y_{k}^{\mathrm{IR}}\hspace*{-2mm}&=&\hspace*{-2mm}\underbrace{\mathbf{h}_k^H\mathbf{x}_k}_{\mbox{desired signal}}\hspace*{-1mm}+\hspace*{-1mm}\underbrace{\sum_{j\ne k}^K\mathbf{h}_k^H\mathbf{x}_j}_{\mbox{multiple-access interference}}+ n^{\mathrm{IR}}_{k},\,\,\\
y_{m}^{\mathrm{ER}}\hspace*{-2mm}&=&\hspace*{-2mm}\mathbf{g}_m^H\sum_{k=1}^K\mathbf{x}_k+n^{\mathrm{ER}}_{m},\,\,
\end{eqnarray}
where $\mathbf{x}_k\in\mathbb{C}^{N_{\mathrm{T}}L\times1}$ denotes the joint transmit signal vector of the $L$ RRHs to IR $k$. The channel between the $L$ RRHs  and  IR $k$ is denoted by $\mathbf{h}_k\in\mathbb{C}^{N_{\mathrm{T}}L\times1}$, and we use $\mathbf{g}_m\in\mathbb{C}^{N_{\mathrm{T}}L\times1}$ to represent the channel  between the $L$ RRHs  and   ER  $m$. We note that the channel vector captures the joint effects of multipath fading and path loss. $n^{\mathrm{IR}}_{k}\sim{\cal CN}(0,\sigma_{\mathrm{s}}^2)$ and $n^{\mathrm{ER}}_{m}\sim{\cal CN}(0,\sigma_{\mathrm{s}}^2)$  are additive white Gaussian noises (AWGN). We assume that the noise variances, $\sigma_{\mathrm{s}}^2$, are identical at all receivers.

\subsection{Signal Model and Backhaul Model}
In each scheduling time slot, $K$ independent signal streams are transmitted simultaneously to the $K$ IRs. Specifically,  a dedicated beamforming vector, $\mathbf{w}_k^l\in\mathbb{C}^{N_{\mathrm{T}}\times1}$, is allocated to IR $k$ at RRH $l\in\{1,\ldots,L\}$ to facilitate information transmission. For  the sake of presentation, we define a  super-vector $\mathbf{w}_k\in\mathbb{C}^{N_{\mathrm{T}}L\times 1}$  for IR $k$ as
\begin{eqnarray}
\mathbf{w}_k=\vect\big([\mathbf{w}_k^1 \,\mathbf{w}_k^2\,\ldots\,\mathbf{w}_k^L]\big).
\end{eqnarray}
  $\mathbf{w}_k$ represents a joint beamformer used by the $L$ RRHs for serving IR $k$. Then, $\mathbf{x}_k$ can be expressed as
\begin{eqnarray}
\mathbf{x}_k=\mathbf{w}_ks_k,
\end{eqnarray}
where $s_k\in\mathbb{C}$ is the data symbol for IR $k$ and  ${\cal E}\{\abs{s_k}^2\}=1,\forall k\in\{1,\ldots,K\}$, is assumed without loss of generality.

On the other hand, the data of each IR is delivered from the CP to the RRHs via backhaul links. The backhaul capacity consumption for backhaul link $l\in\{1,\ldots,L\}$ is given by
\begin{eqnarray}
C^{\mathrm{Backhaul}}_l=\sum_{k=1}^K\bnorm{\Big[\norm{\mathbf{w}_k^l}_2\Big]}_0\,\, R_k,
\end{eqnarray}
where $R_k$ is the required backhaul data rate for conveying the data of IR $k$ to a RRH. We note that the backhaul links  may be capacity-constrained and the CP may not be able to send the data to all RRHs as required for  full cooperation. Thus, to reduce the load on the backhaul links, the CP can enable partial cooperation by sending the data of information  receiver $k$  only to a subset of  the RRHs. In particular, by setting $\mathbf{w}_l^k=\mathbf{0}$, RRH $l$ is not participating in the joint data transmission to IR $k$.  Thus, the CP is not required to send the data for IR $k$ to  RRH $l$  via the backhaul link which leads to a lower backhaul link capacity consumption.
\subsection{RRH Power Supply Model}
In the considered CoMP network, we assume that the CP transfers energy to the RRHs  for supporting the power consumption at the RRHs and
facilitating a  more efficient network operation. In particular,
$ E^{\mathrm{s}}_l- (E^{\mathrm{s}}_l)^2\beta_l$ units of energy are transferred to RRH $l$ via a dedicated power line where
 $E^{\mathrm{s}}_l,\forall l\in\{1,\ldots,L\}$, is the power supplied by the CP to RRH $l$.  $(E^{\mathrm{s}}_l)^2\beta_l$ is the power loss in delivering the power from the CP RRH $l$. $\beta>0$ is a constant proportional to the ratio between the resistance of the adopted power line and the voltage of power transmission. We note that  $E^{\mathrm{s}}_l- (E^{\mathrm{s}}_l)^2\beta_l\ge 0$ always hold by the law of conservation of energy.
\section{Problem Formulation}
\label{sect:forumlation}

\subsection{Achievable Rate and Energy Harvesting}
\label{subsect:Instaneous_Mutual_information}
The  achievable rate (bit/s/Hz) between the $L$ RRHs and IR $k$ is given by
\begin{eqnarray}
C_{k}=\log_2(1+\Gamma_{k}),\,\,
\mbox{where}\quad
\Gamma_{k}=\frac{\abs{\mathbf{h}_k^H\mathbf{w}_k}^2}{\sum\limits_
{\substack{j\neq k}}^K\abs{\mathbf{h}_k^H\mathbf{w}_j}^2+\sigma_{\mathrm{s}}^2}
\end{eqnarray}
is the receive signal-to-interference-plus-noise ratio (SINR) at IR $k$.

On the other hand, the information signal, $\mathbf{w}_k s_k,\forall k\in\{1,\ldots,K\}$,  serves as a dual purpose carrier for conveying both information and energy concurrently  in the considered system. The total amount of energy\footnote{We adopt the normalized energy unit  Joule-per-second in this paper.  Therefore,
the terms ``power" and ``energy" are used interchangeably.} harvested by ER $m\in\{1,\ldots,M\}$ is given by
\begin{eqnarray}\label{eqn:ER_power}
E_{m}^{\mathrm{ER}}=\mu\Big(\sum_{k=1}^K\abs{\mathbf{g}_m^H\mathbf{w}_k}^2\Big),
\end{eqnarray}
where $0<\mu\leq1$ denotes the efficiency of the conversion of the received RF energy to electrical energy for storage. We assume that $\mu$ is a constant and is identical for all ERs. Besides, the contribution of the antenna noise power to the harvested energy  is negligibly small compared to the harvested energy from the information signal, $\abs{\mathbf{g}_m^H\mathbf{w}_k}^2$, and thus is neglected in (\ref{eqn:ER_power}).

\subsection{Optimization Problem Formulation}
\label{sect:cross-Layer_formulation}
The system objective is to jointly minimize the weighted sum of the total network transmit power and the maximum capacity consumption per backhaul link while providing QoS for reliable communication and power transfer. The resource allocation algorithm design  is formulated as the following optimization problem:\vspace*{-2mm}
\begin{eqnarray} \label{eqn:cross-layer}\notag
&&\hspace*{-10mm} \underset{ E^{\mathrm{s}}_l,\mathbf{w}_k}{\mino}\,\, \delta\max_{l\in\{1,\ldots,L\}}\Big\{C^{\mathrm{Backhaul}}_l \Big\}+\eta \sum_{k=1}^K\sum_{l=1}^L\norm{\mathbf{w}^l_k}_2^2\\
\notag \mbox{s.t.}\hspace*{-2mm} &&\hspace*{2mm}\mbox{C1: }\Gamma_{k}\ge\Gamma_{\mathrm{req}_k},\,\, \forall k, \notag\\
&&\hspace*{2mm}\mbox{C2: }P_{\mathrm{C}}^\mathrm{CP}+\sum_{l=1}^L E^{\mathrm{s}}_l\le P_{\max}^{\mathrm{CP}},\notag\\
&&\hspace*{2mm}\mbox{C3: }P_{\mathrm{C}_l}+ \varepsilon\sum_{k=1}^K\norm{\mathbf{w}^l_k}^2_2\le E^{\mathrm{s}}_l- (E^{\mathrm{s}}_l)^2\beta_l,\,\, \forall l,\notag\\
&&\hspace*{2mm}\mbox{C4: }\sum_{k=1}^K\norm{\mathbf{w}^l_k}^2_2\le P^{\mathrm{T}_{\max}}_l,\,\, \forall l,\notag\\
&&\hspace*{-2mm}\mbox{C5:}\,\, E_{m}^{\mathrm{ER}}\ge P^{\min}_{m},\,\, \forall m,\,\, \,\, \,\,
\mbox{C6:}\,\,  E^{\mathrm{s}}_l\ge 0,\,\, \forall l,
\end{eqnarray}
where $\delta\ge0$ and $\eta\ge0$ in the objective function are constants which  reflect the preference of the system operator for  the capacity consumption of individual backhaul links and the total network transmit power consumption, respectively. Besides, $\delta$ can also be interpreted as the energy/power cost in conveying information to the RRHs via backhaul.  $\Gamma_{\mathrm{req}_k}>0$ in constraint C1 indicates the required minimum  receive SINR at IR $k$ for information decoding.  The corresponding data rate per backhaul link use for IR $k$ is given by $R_k=\log_2(1+\Gamma_{\mathrm{req}_k})$. In C2, $P_{\mathrm{C}}^\mathrm{CP}$ and  $P_{\max}^{\mathrm{CP}}$ are the hardware circuit power consumption and  the maximum  power available at the CP, respectively. In C3, $P_{\mathrm{C}_l}$ and $ E^{\mathrm{s}}_l- (E^{\mathrm{s}}_l)^2\beta_l\ge0$ are the hardware circuit power consumption and the maximum available power at RRH $l$, respectively. $\varepsilon\ge 1$  is a constant which accounts for the power inefficiency of the power amplifier.
$P^{T_{\max}}_l$ in C4 is the maximum transmit power allowance for RRH $l$, which can be used to limit out-of-cell
interference. Constant $P^{\min}_{m}$  in constraint C5 specifies the required minimum  harvested energy  at ER $m$.  C6 is the non-negativity constraint on the power optimization variables.
\begin{Remark} We note that the objective function considered in this paper is different from that in \cite{JR:Quek} and \cite{CN:Wei_yu_sparse_BF}. In particular, we focus on the capacity consumption of  individual backhaul links while \cite{JR:Quek} and \cite{CN:Wei_yu_sparse_BF} studied the total network backhaul capacity consumption. Although the considered problem formulation does not constrain the capacity consumption of the individual backhaul links, it provides a first-order measure of the backhaul loading in the considered CoMP network when enabling partial cooperation. This information provides system design insight for the required backhaul deployment.
\end{Remark}
\section{Resource Allocation Algorithm Design}
The optimization problem in (\ref{eqn:cross-layer}) is a non-convex  problem due to the non-convexity of the objective function,
constraint C1, and constraint C5. In particular, the combinatorial nature of the objective function  results in an NP-hard  optimization problem \cite{JR:Quek}. To strike a balance between system performance and computational complexity, we develop an iterative algorithm  for obtaining a suboptimal solution.  To this end, we first reformulate the optimization problem by approximating the original non-convex objective function as a weighted sum of convex functions with different weight factors. Then, we recast the reformulated problem as a semidefinite programming (SDP) problem via SDP relaxation and solve it optimally. Subsequently, a suboptimal solution to the original optimization problem is obtained by updating the weight factors and solving the reformulated problem iteratively.
\subsection{Convex Relaxation}
The non-convex weighted capacity consumption of backhaul link $l$, $\delta C^{\mathrm{Backhaul}}_l$, can be approximated as follows:
\begin{eqnarray}\label{eqn:objective_approx}
\delta C^{\mathrm{Backhaul}}_l  \hspace*{-2mm}&\stackrel{(a)}{=}&\hspace*{-2mm}\delta \sum_{k=1}^K\bnorm{\Big[\norm{\mathbf{w}_k^l}_2^2\Big]}_0\,\, R_k\\
\hspace*{-2mm}&\stackrel{(b)}{\approx}&\hspace*{-2mm}\delta\sum_{k=1}^K\bnorm{\Big[\rho_k^l\norm{\mathbf{w}_k^l}_2^2\Big]}_1 R_k=\delta\sum_{k=1}^K \rho_k^l\norm{\mathbf{w}_k^l}_2^2 R_k\notag
\end{eqnarray}
where $\rho_k^l\ge 0,\forall k,l,$ in $(b)$ are given constant weight factors which can be used to achieve solution sparsity. $(a)$ indicates that the value of the $l_0$-norm  is invariant when the input arguments are squared. $(b)$ is due to the fact that the $l_0$-norm can be approximated by its  convex hull which is the $l_1$-norm.  This approximation is known as convex relaxation and is commonly used in the field of compressed sensing for handling $l_0$-norm optimization problems \cite{JR:Quek}--\nocite{CN:Wei_yu_sparse_BF,JR:compressed_sensing}\cite{JR:compressive_sensing_boyd}.

\subsection{SDP Relaxation}
We substitute  (\ref{eqn:objective_approx}) into (\ref{eqn:cross-layer}) and define $\mathbf{W}_k=\mathbf{w}_k\mathbf{w}_k^H$, $\mathbf{H}_k=\mathbf{h}_k\mathbf{h}_k^H$, and $\mathbf{G}_m=\mathbf{g}_m\mathbf{g}_m^H$. Then, we recast the reformulated problem in its epigraph form \cite{book:convex} which is given as follows:
\begin{eqnarray}\label{eqn:rank_one}
&&\hspace*{-10mm} \underset{\mathbf{W}_k\in \mathbb{H}^{N_{\mathrm{T}}},E^{\mathrm{s}}_l,\phi}{\mino}\,\, \phi+ \eta\sum_{k=1}^K\Tr(\mathbf{W}_k)\notag\\
\mbox{s.t.} &&\hspace*{-5mm}\mbox{C1: }\frac{\Tr(\mathbf{H}_k\mathbf{W}_k)}{\Gamma_{\mathrm{req}_k}}\ge\sum\limits_
{\substack{j\neq k}}^K\Tr(\mathbf{H}_k\mathbf{W}_j)+\sigma_{\mathrm{s}}^2,\forall k,\notag \\
&&\hspace*{10mm}{\mbox{C2}},\quad{\mbox{C6}},\notag\\
&&\hspace*{-5mm}\mbox{C3: } P_{\mathrm{C}_l}+\varepsilon\sum_{k=1}^K\Tr\big(\mathbf{B}_l\mathbf{W}_k\big)\le E^{\mathrm{s}}_l- (E^{\mathrm{s}}_l)^2\beta_l,\,\, \forall l,\notag\\
&&\hspace*{-5mm}\mbox{C4: } \sum_{k=1}^K\Tr\big(\mathbf{B}_l\mathbf{W}_k\big)\le  P^{\mathrm{T}_{\max}}_l,\,\, \forall l,\notag\\
&&\hspace*{-5mm}\mbox{C5:}\notag\mu\Big(\sum_{k=1}^K\Tr\big(\mathbf{W}_k\mathbf{G}_m\big)\Big)\ge P^{\min}_{m},\,\, \forall m, \notag\\
&&\hspace*{-5mm}\mbox{C7:} \delta \Big(\sum_{k=1}^K\Tr(\mathbf{W}_k\mathbf{B}_l)\rho_k^lR_k \Big)\le \phi,\forall l,\notag\\
&&\hspace*{-15mm}\mbox{C8:}\,\, \mathbf{W}_k\succeq \mathbf{0},\,\, \forall k, \quad\,\,\,
\mbox{C9:}\,\, \Rank(\mathbf{W}_k)\le 1,\,\, \forall k,
\end{eqnarray}
where
\begin{eqnarray}
&&\hspace*{-5mm}\mathbf{B}_l\triangleq\diag\Big(\underbrace{0,\cdots,0}_{(l-1)N_\mathrm{T}},\underbrace{1,\cdots,1}_{N_\mathrm{T}},\underbrace{0,\cdots,0}_{(L-l)N_\mathrm{T}}\Big),\forall l\in\{1,\ldots,L\}\notag,
\end{eqnarray}
is a block diagonal matrix with $\mathbf{B}_l\succeq \mathbf{0}$. $\phi$ in the objective function and constraint C7 is an auxiliary optimization variable.   Constraints C8, C9, and $\mathbf{W}_k\in\mathbb{H}^{N_{\mathrm{T}}},\forall k$, are imposed to guarantee that $\mathbf{W}_k=\mathbf{w}_k\mathbf{w}_k^H$ holds after optimization.

\begin{table}[t]\caption{Iterative Resource Allocation Algorithm}\label{table:algorithm}
\vspace*{-0.6cm}
\renewcommand\thealgorithm{}
\begin{algorithm} [H]                    
\caption{Reweighted $l_1$-norm Method}          
\label{alg1}                           
\begin{algorithmic} [1]
\normalsize        
\STATE Initialize the maximum number of iterations $L_{\max}$ and a small constant $\kappa\rightarrow 0$
\STATE Set iteration index $n=0$ and $\rho_{k}^l(n)=1,\forall k,l$

\REPEAT [Loop]
\STATE Solve  (\ref{eqn:sdp_relaxation})  for a given set of $\rho_{k}^l(n)$ and obtain an intermediate beamforming vector $\mathbf{w}_k^l$
\STATE Update the weight factor as follows:
\begin{eqnarray}
\rho_{k}^l (n+1)&=&\frac{1}{\norm{\mathbf{w}_k^l}_2^2+\kappa}, \forall l,k,\notag\\
n&=&n+1\notag
\end{eqnarray}

\UNTIL{ $n=L_{\max}$}

\end{algorithmic}
\end{algorithm}
\vspace*{-0.9cm}\normalsize
\end{table}

Then, we relax constraint $\mbox{C9: }\Rank(\mathbf{W}_k)\le1$ by removing it from the problem formulation, such that the considered problem becomes a convex SDP  given by
\begin{eqnarray}
\label{eqn:sdp_relaxation}&&\hspace*{-25mm} \underset{\mathbf{W}_k\in \mathbb{H}^{N_{\mathrm{T}}},E^{\mathrm{s}}_l,\phi}{\mino}\,\, \phi+ \eta\sum_{k=1}^K\Tr(\mathbf{W}_k)\notag\\
\hspace*{2mm}\mbox{s.t.} &&\hspace*{-3mm}\mbox{C1 -- C8}.
\end{eqnarray}
We note that the relaxed problem in (\ref{eqn:sdp_relaxation}) can be solved efficiently by numerical solvers such as  CVX \cite{website:CVX}. If the solution $\mathbf{W}_k$ of (\ref{eqn:sdp_relaxation}) is a rank-one matrix, then  the problems in (\ref{eqn:rank_one}) and (\ref{eqn:sdp_relaxation}) share the same optimal solution and the same optimal objective value. Otherwise, the optimal objective value of (\ref{eqn:sdp_relaxation}) serves as a lower bound for the objective value of (\ref{eqn:rank_one}).

Next, we  reveal the tightness of the SDP relaxation adopted in  (\ref{eqn:sdp_relaxation}) in the following theorem.
\begin{Thm}\label{thm:rankone_condition} Assuming the channel vectors of the IRs, $\mathbf{h}_k,k\in\{1,\ldots,K\},$ and the ERs, $\mathbf{g}_m,m\in\{1,\ldots,M\},$ can be modeled as statistically independent  random variables then the solution of (\ref{eqn:sdp_relaxation}) is rank-one, i.e.,  $\Rank(\mathbf{W}_k)=1,\forall k$, with probability one.
\end{Thm}
\emph{\quad Proof: } Please refer to the Appendix. \qed

In other words,  whenever  the channels satisfy the condition stated in Theorem 1, the optimal  beamformer $\mathbf{w}^*_k$ of (\ref{eqn:rank_one}) can be obtained with probability one by performing an eigenvalue decomposition of the solution $\mathbf{W}_k$ of (\ref{eqn:sdp_relaxation}) and selecting the principal eigenvector as the beamformer.

\subsection{Iterative Resource Allocation Algorithm}
In general, for a fixed weight factor, $\rho_k^l$, the solution of (\ref{eqn:rank_one}) does not necessarily provide sparsity and the approximation adopted in (\ref{eqn:objective_approx}) may not be tight. For improving the obtained solution, we adopt the \emph{Reweighted $l_1$-norm Method} which was originally designed to enhance the data acquisition in compressive sensing \cite{JR:compressive_sensing_boyd}. The overall resource allocation algorithm is summarized in Table \ref{table:algorithm}. In particular, the weight factor $\rho_k^l$ is updated as  in line 5 of the iterative algorithm such that the magnitude of beamforming vectors $\norm{\mathbf{w}_k^l}_2^2$ with small values are further reduced in the next iteration. As a result, by iteratively updating $\rho_k^l$ and solving (\ref{eqn:sdp_relaxation}), a suboptimal beamforming solution with sparsity can be constructed.  We note that the  iterative algorithm in Table \ref{table:algorithm} converges to a local optimal solution of the original problem formulation in (\ref{eqn:cross-layer}) for $\kappa\rightarrow 0$ and a sufficient number of iterations \cite{JR:Quek,JR:compressive_sensing_boyd}. Furthermore, when the primal-dual path-following
method   \cite{JR:SDP_relaxation1} is used  by the  numerical solver for solving (\ref{eqn:sdp_relaxation}),  the computational complexity of the proposed algorithm is $\bigo(L_{\max}\max\{N_{\mathrm{T}}L,K+3L+M\}^4(N_{\mathrm{T}}L)^{1/2}\log(1/\epsilon))$ for a given solution accuracy $\epsilon>0$. The computational complexity  is significantly reduced compared to the computational complexity of an exhaustive search with respect to $K$ and $L$, i.e.,  $\bigo((2^L-1)^K\max\{N_{\mathrm{T}}L,K+3L+M\}^4(N_{\mathrm{T}}L)^{1/2}\log(1/\epsilon))$.

\begin{figure}[t]
\centering
\includegraphics[width=2.0in]{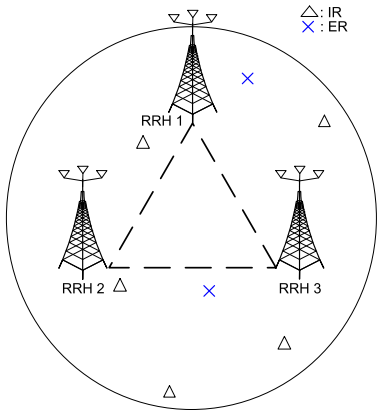}\vspace*{-2mm}
\caption{CoMP network simulation topology with $L=3$ RRHs, $K=5$ IRs, and $M=2$ ERs.}
\label{fig:simulation_model}
\end{figure}

\begin{table}[t]\caption{System parameters}\label{tab:feedback} \centering
\begin{tabular}{|L|l|}\hline
\hspace*{-1mm}Carrier center frequency & 1.9 GHz  \\
\hline
\hspace*{-1mm}Path loss exponent &  3.6  \\
\hline
\hspace*{-1mm}Multipath fading distribution & \mbox{Rayleigh fading}  \\

\hline
\hspace*{-1mm}Total noise variance, $\sigma_{\mathrm{s}}^2$ &  \mbox{$-23$ dBm}   \\
\hline
\hspace*{-1mm}Minimum required SINR, \hspace*{-1mm}$\Gamma_{\mathrm{req}_k}=\Gamma_{\mathrm{req}},\forall k\in\{1,\ldots,K\}$ &  \mbox{$15$ dB}   \\
\hline
\hspace*{-1mm}Circuit power consumption at CP, $P_{\mathrm{CP}}^{\mathrm{C}}$ &  $40$ \mbox{dBm}   \\
\hline
\hspace*{-1mm}Circuit power consumption at the $l$-th RRH, $P_{\mathrm{C}_l}$ &  $30$ \mbox{dBm}   \\
\hline
\hspace*{-1mm}Max. power  supply at the CP, $P^{\mathrm{CP}}_{\max}$ &  $50$ \mbox{dBm}   \\
\hline
\hspace*{-1mm}Power amplifier power efficiency  & $1/{\varepsilon}=0.38$  \\
\hline
\hspace*{-1mm}Max. transmit power  allowance, $P_l^{T_{\max}}$ & $46$ dBm \\
\hline
\hspace*{-1mm}Min. required power transfer, $P_{m}^{\min}$ & $0$ dBm  \\
\hline
\hspace*{-1mm}RF to electrical energy conversion efficiency, $\mu$ & $0.5$  \\
\hline
\hspace*{-1mm}Power loss in transferring power from the CP to \hspace*{-1mm}RRH, $1-\beta_l$ &  $0.2$   \\
           \hline

\end{tabular}\vspace*{-4mm}
\end{table}

\section{Results}
In this section, we evaluate  the network performance of the proposed resource allocation design via simulations.
 There are $L=3$ RRHs, $K=5$ IRs, and $M=2$ ERs in the system.  We focus on the network topology  shown in Figure \ref{fig:simulation_model}. The distance between any two RRHs is $500$ meters. The three RRHs construct an equilateral triangle while the IRs and ERs are uniformly distributed inside a disc with  radius  $1000$ meters centered at the centroid of the triangle.   The simulation parameters can be found in Table \ref{tab:feedback}. In the iterative algorithm, we set $\kappa$ and $L_{\max}$ to $0.0001$ and $20$, respectively.  The numerical results in this section were
averaged over 1000 independent channel realizations for both
path loss and multipath fading.  The performance of the proposed scheme is compared with the performances of a full cooperation scheme, an optimal exhaustive search scheme, and a traditional system with co-located transmit antennas. For the full cooperation scheme, the solution is obtained by setting $\delta=0$, $\eta=1$, and solving (\ref{eqn:sdp_relaxation}) by SDP relaxation. For the exhaustive search, it is expected that multiple optimal solutions for (\ref{eqn:cross-layer}) may exist. Thus, for the set of  optimal solutions, we further select the one having the minimal total system backhaul capacity consumption. If there are  multiple optimal solutions with the same total system backhaul capacity consumption, then we select the one requiring the minimal  total  network transmit power.    As for the co-located transmit antenna system, we assume that there is only one RRH located  at the center of the system equipped with the same number of antennas as all RRHs combined in the distributed stetting. Besides, the CP is not at the same location as the  RRH for the  co-located transmit antenna system, i.e., a backhaul is still needed. Furthermore, we set $P_l^{T_{\max}}=\infty$ for the co-located transmit antenna system to study its power consumption.



\begin{figure}[t]
\subfigure[Average maximum capacity consumption per backhaul link.]{
\includegraphics[width=3.5 in]{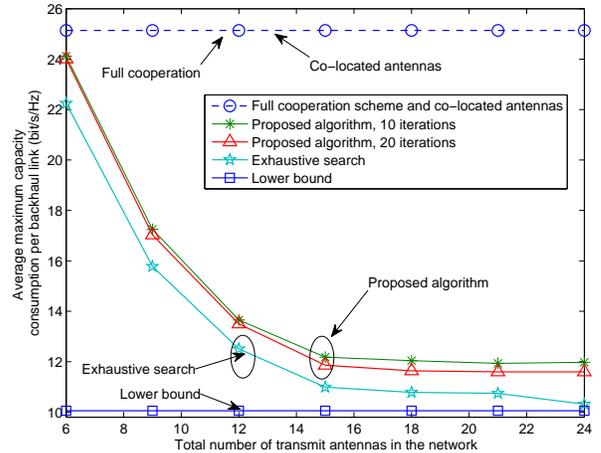}
\label{fig:backhaul_nt_subfig1} }\vspace*{-1mm}
\subfigure[Average  total system backhaul capacity consumption.]{
\includegraphics[width=3.5 in]{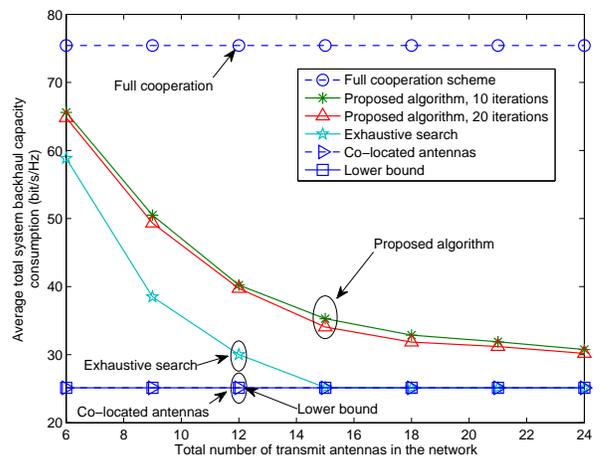}
\label{fig:backhaul_nt_subfig2} }\caption[Optional caption for list of figures]
{Average backhaul capacity consumption versus   total number of transmit antennas in the network for  different resource allocation schemes.}
\label{fig:backhaul_nt}
\end{figure}


\subsection{Average Backhaul Capacity Consumption}
In Figures \ref{fig:backhaul_nt_subfig1} and \ref{fig:backhaul_nt_subfig2}, we study the average maximum backhaul capacity consumption per backhaul link and the average total system backhaul capacity consumption, respectively, versus  the total number of transmit antennas in the network,  for different resource allocation schemes. We set $\delta=1$ and $\eta=0$ in (\ref{eqn:sdp_relaxation}) for the proposed scheme
to fully minimize the maximum capacity consumption per backhaul link. The performance of the proposed  iterative algorithm is shown for $10$ and $20$ iterations. It can be seen from Figure \ref{fig:backhaul_nt_subfig1} that the proposed iterative algorithm achieves a close-to-optimal backhaul capacity consumption in all considered scenarios even for the case of $10$ iterations. We note that the gap between the proposed algorithm and the exhaustive search in Figure \ref{fig:backhaul_nt_subfig1}  is caused by the sub-optimality of
the objective function approximation in (\ref{eqn:objective_approx}) and insufficient numbers of
iterations. In fact, the superior average maximum system backhaul capacity consumption of the optimal exhaustive scheme in Figure  \ref{fig:backhaul_nt_subfig1}   compared to the proposed scheme comes at the expense of an exponential  computational complexity with respect to the number of IRs and RRHs.  On the other hand, the performance gap between the proposed iterative resource allocation algorithm and the full cooperation scheme(/co-located antennas system) increases  as the total number of transmit antennas. In Figures \ref{fig:backhaul_nt_subfig1} and \ref{fig:backhaul_nt_subfig2}, we observe that the average backhaul capacity consumption of the proposed algorithm decreases monotonically with an increasing number of antennas and converges to constant values close to the lower bounds, respectively. The lower bounds in Figures  \ref{fig:backhaul_nt_subfig1} and \ref{fig:backhaul_nt_subfig2} are given by $\ceil[\Big]{\frac{K}{L}}\log_2(1+\Gamma_{\mathrm{req}})$ and $K\log_2(1+\Gamma_{\mathrm{req}})$, respectively.  Indeed, when both the power budget and the number of antennas at the RRHs are sufficiently large, full cooperation may not be beneficial. In this case, conveying the data of each IR to one RRH may be sufficient for providing the QoS requirements for reliable communication and efficient power transfer. Hence, backaul system resources can be saved. Besides, it can be seen from  Figure \ref{fig:backhaul_nt_subfig2} that the system with co-located antennas requires the smallest amount of total system backhaul capacity since the data of each IR is conveyed  only to a single RRH. However, the superior performance of the co-located antenna system in terms of total network backhaul capacity consumption incurs the highest capacity consumption per backhaul link among all the schemes, cf. Figure \ref{fig:backhaul_nt_subfig1}.

\subsection{Average Total Transmit Power and Harvested Power}
\begin{figure}[t]
\centering
\includegraphics[width=3.5in]{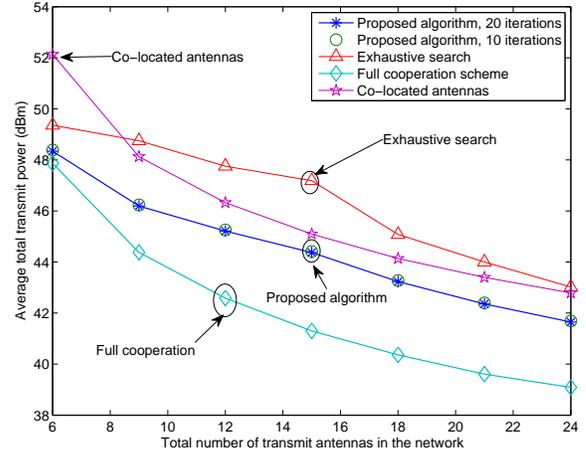}\vspace*{-2mm}
\caption{Average total transmit power (dBm)  versus total number of transmit antennas in the network for different  resource allocation schemes. }
\label{fig:PT_NT}
\end{figure}

In Figure \ref{fig:PT_NT}, we study the average total transmit power  versus total number of transmit antennas for  different resource allocation schemes.  It can be observed that the total transmit power decreases monotonically with increasing  number of transmit antennas. This is due to the fact that the degrees of freedom for resource allocation increase with the number of transmit antennas, which enables a more power efficient resource allocation. Besides, the proposed algorithm consumes a lower transmit power compared to the optimal exhaustive search scheme. This is because the exhaustive search scheme consumes a smaller backhaul capacity at the expense of a higher transmit power. Furthermore, the system with co-located antennas consumes a  higher transmit power than the proposed scheme and the full cooperation scheme in all considered scenarios which reveals the power saving potential of CoMP due to its inherent spatial diversity.  On the other hand, it is expected that the full cooperation scheme is able to achieve the lowest average total transmit power at the expense of an exceedingly large backhaul capacity consumption, cf. Figure \ref{fig:backhaul_nt}.

\begin{figure}[t]
\centering
\includegraphics[width=3.5in]{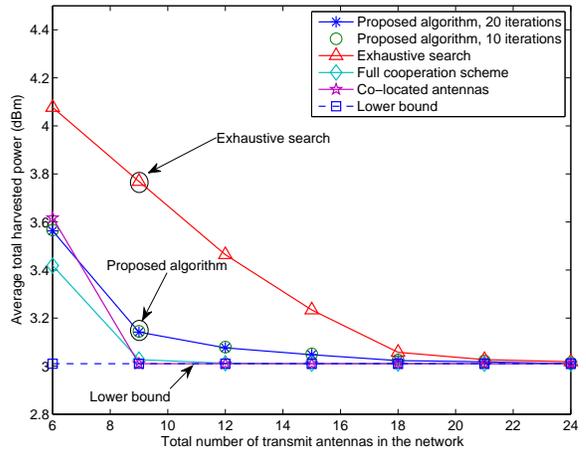}\vspace*{-2mm}
\caption{Average total harvested power (dBm) versus total number of transmit antennas in the network for different  resource allocation schemes. }
\label{fig:hp_NT}
\end{figure}

In Figure \ref{fig:hp_NT}, we study
the average total harvested power versus the total number of transmit antennas in the network for  different resource allocation schemes.  We compare the average total harvested power of all resource allocation schemes with a lower  bound which is computed  by assuming that constraint C5 is satisfied  with equality for all ERs.  As can be observed, the total
average harvested powers in all considered scenarios are monotonically non-increasing with respect to the number of transmit antennas.
This is because the extra degrees of freedom offered by the increasing number of antennas improve the efficiency of resource allocation. In particular,  the direction of beamforming
matrix $\mathbf{W}_k$ can be more accurately steered towards the IRs which reduces the power allocation to $\mathbf{W}_k$ and the leakage of power to the ERs.  This also explains the lower harvested power for the full cooperation scheme and the system with co-located antennas since they both exploit all transmit antennas in the network for joint transmission. On the other hand, the highest amount of radiated power can be harvested for the exhaustive search scheme at the expense of a higher total transmit power.

\section{Conclusions}\label{sect:conclusion}
In this paper, we studied the  resource allocation algorithm design for CoMP multiuser communication systems with SWIPT.  The algorithm design was formulated as a non-convex combinatorial  optimization problem with the objective to jointly minimize the total network transmit power and the maximum capacity consumption of the backhaul links. The proposed problem formulation took into account QoS requirements for communication reliability and power transfer.  A  suboptimal iterative resource allocation algorithm was proposed for obtaining a locally optimal solution of the considered problem.  Simulation results
showed that the proposed suboptimal iterative resource allocation scheme  performs close to the optimal  exhaustive search scheme and provides a substantial reduction in backhaul capacity consumption  compared to  full cooperation. Besides, our results unveiled the potential power savings enabled by CoMP networks compared to centralized systems with multiple antennas co-located  for SWIPT.

\section*{Appendix-Proof of Theorem \ref{thm:rankone_condition}}
It can be verified that (\ref{eqn:sdp_relaxation}) satisfies Slater's constraint qualification and is jointly convex with respect to the optimization variables. Thus, strong duality holds and solving the dual problem  is equivalent to solving the primal problem \cite{book:convex}.  For the dual problem, we need the Lagrangian function of the primal problem in (\ref{eqn:sdp_relaxation}) which is given by
\begin{eqnarray}
&&\hspace*{-6mm}{\cal L}\Big(\mathbf{W}_k,E^{\mathrm{s}}_l,\phi,\mathbf{Y}_k,\psi_l,\xi_k,\tau_m,\lambda,\omega_l,\theta_l,\chi_l\Big)\\
\hspace*{-2mm}=&&\hspace*{-6mm}\sum_{k=1}^K\Tr(\mathbf{A}_k\mathbf{W}_k)-\sum_{k=1}^K\Tr\Big(\mathbf{W}_k
\big(\mathbf{Y}_k+\frac{\xi_k\mathbf{H}_k}{\Gamma_{\mathrm{req}_k}}\big)\Big)+\Delta\notag
\end{eqnarray}
where\begin{eqnarray}\label{eqn:A_k}
\hspace*{-2mm}\mathbf{A}_k\hspace*{-3mm}&=&\hspace*{-3mm}\mathbf{D}_k+\sum_{j\neq k}^K\xi_j\mathbf{H}_j-\mu\sum_{m=1}^M\tau_m\mathbf{G}_m, \\
\hspace*{-2mm}\mathbf{D}_k\hspace*{-3mm}&=&\hspace*{-3mm} R_k\delta\sum_{l=1}^{L}\mathbf{B}_l\rho_k^l\chi_l +\eta\mathbf{I}_{N_{\mathrm{T}}}+
\sum_{l=1}^L(\psi_l+\theta_l)\varepsilon\mathbf{B}_l,\, \mbox{and}\\
\hspace*{-2mm}\Delta\hspace*{-3mm}&=&\hspace*{-3mm}\phi+\lambda(P_{\mathrm{C}}^\mathrm{CP}\hspace*{-0.5mm}+\hspace*{-0.5mm}\sum_{l=1}^L E^{\mathrm{s}}_l\hspace*{-0.5mm}- \hspace*{-0.5mm}P_{\max}^{\mathrm{CP}})\hspace*{-0.5mm}+\hspace*{-0.5mm}\sum_{m=1}^M \tau_m P^{\min}_m\hspace*{-0.5mm}-\hspace*{-0.5mm} \sum_{l=1}^L \omega_lE^{\mathrm{s}}_l\notag\\
&&\hspace*{-12mm}+\sum_{k=1}^K\xi_k\sigma_{\mathrm{s}}^2+\sum_{l=1}^L \Big[\psi_l(P_{\mathrm{C}_l}\hspace*{-0.5mm}-\hspace*{-0.5mm} ( E^{\mathrm{s}}_l- ( E^{\mathrm{s}}_l)^2\beta_l))\hspace*{-0.5mm}-\hspace*{-0.5mm}\theta_lP^{\mathrm{T}_{\max}}_l\hspace*{-0.5mm}-\hspace*{-0.5mm}\chi_l\phi\Big].\notag
\end{eqnarray}
Here, $\Delta$ denotes the collection of terms that only involve variables that are independent of $\mathbf{W}_k$. $\mathbf{Y}_{k}$ is the dual variable matrix for constraint C8. $\xi_k$, $\lambda$, $\psi_l$, $\theta_l$,  $\tau_m$, $\omega_l$, and $\chi_l$ are the scalar dual variables for constraints C1--C7, respectively.

Then, the dual problem of (\ref{eqn:sdp_relaxation}) is given by\small
\begin{equation}\label{eqn:dual}
\underset{\underset{\theta_l,\lambda,\omega_l\ge 0,\mathbf{Y}_k\succeq \mathbf{0}}{\psi_l,\xi_k,\tau_m,\chi_l\ge 0}}{\maxo} \,\underset{\underset{E^{\mathrm{s}}_l,\phi}{\mathbf{W}_k\in\mathbb{H}^{N_{\mathrm{T}}}}}{\mino} \,\,{\cal L} \Big(\hspace*{-0.5mm}\mathbf{W}_k,E^{\mathrm{s}}_l,\phi,\hspace*{-0.5mm}\mathbf{Y}_k,\psi_l,
\xi_k,\tau_m,\lambda,\omega_l,\theta_l,\chi_l\hspace*{-0.5mm}\Big)
\end{equation}\normalsize
subject to $\sum_{l=1}^K \chi_l=1$. For the sake of notational simplicity, we define $\mathbf{\Upsilon}^*\triangleq\{\mathbf{W}_k^*,E^{\mathrm{s}*}_l,\phi^*\}$   and $\mathbf{\Xi}^*\triangleq\{\mathbf{Y}_k^*,\psi_l^*,\xi_k^*,\tau_m^*,\lambda^*,\omega_l^*,\theta_l^*,\chi_l^*\}$ as the set of optimal primal and  dual variables of  (\ref{eqn:sdp_relaxation}), respectively. Now, we consider the following Karush-Kuhn-Tucker (KKT) conditions which are useful in the proof:
\begin{eqnarray}
\hspace*{-3mm}\mathbf{Y}_k^*\hspace*{-3mm}&\succeq&\hspace*{-3mm}\mathbf{0},\,\,\tau_m^*,\,\psi_l^*,\xi_k^*\ge 0,\,\forall k,\,\forall m,\,\forall l, \label{eqn:dual_variables}\\
\hspace*{-3mm}\mathbf{Y}_k^*\mathbf{W}_k^*\hspace*{-3mm}&=&\hspace*{-3mm}\mathbf{0},\label{eqn:KKT-complementarity}\\
\hspace*{-3mm}\mathbf{Y}_k^*\hspace*{-3mm}&=&\hspace*{-3mm}\mathbf{A}_{k}^*-\xi_k^*\frac{\mathbf{H}_k}{\Gamma_{\mathrm{req}_k}},
\label{eqn:lagrangian_gradient}
\end{eqnarray}
where $\mathbf{A}_{k}^*$ is obtained by substituting the optimal dual variables $\mathbf{\Xi}^*$ into (\ref{eqn:A_k}). $\mathbf{Y}_k^*\mathbf{W}_k^*=\mathbf{0}$ in (\ref{eqn:KKT-complementarity}) indicates that for $\mathbf{W}^*_k\ne\mathbf{0}$, the columns of $\mathbf{W}^*_k$ are in the null space of $\mathbf{Y}^*_k$. Therefore, if  $\Rank(\mathbf{Y}^*_k)=N_{\mathrm{T}}L-1$, then the optimal beamforming matrix $\mathbf{W}^*_k\ne \mathbf{0}$ must be a rank-one matrix.  We now show by contradiction that $\mathbf{A}_k^*$ is a positive definite matrix with probability one in order to reveal the structure of $\mathbf{Y}^*_k$. Let us focus on the dual problem in (\ref{eqn:dual}). For a given set of optimal dual variables, $\mathbf{\Xi}^*$ ,  power supply variables, $E^{\mathrm{s}*}_l$, and auxiliary variable $\phi^*$,  the dual problem in (\ref{eqn:dual}) can be written as
\begin{eqnarray}\hspace*{-2mm}\label{eqn:dual2}
\,\,\underset{\mathbf{W}_k\in\mathbb{H}^{N_{\mathrm{T}}}}{\mino} \,\, {\cal L}\Big(\hspace*{-0.5mm}\mathbf{W}_k,\phi^*,E^{\mathrm{s}*}_l,\mathbf{Y}_k^*,\psi_l^*,\xi_k^*,\tau_m^*,\lambda^*,\omega_l^*,\theta_l^*,\chi_l^*\hspace*{-0.5mm}\Big).
\end{eqnarray}
Suppose $\mathbf{A}_k^*$ is not positive definite, then we can choose $\mathbf{W}_k=r\mathbf{w}_k\mathbf{w}_k^H$ as one of the optimal solutions of (\ref{eqn:dual2}), where $r>0$ is a scaling parameter and $\mathbf{w}_k$ is the eigenvector corresponding to one of the non-positive eigenvalues of $\mathbf{A}_k^*$.  We substitute $\mathbf{W}_k=r\mathbf{w}_k\mathbf{w}_k^H$ into (\ref{eqn:dual2}) which leads to
\begin{equation}
\underbrace{\sum_{k=1}^K\Tr(r\mathbf{A}_k^*\mathbf{w}_k\mathbf{w}_k^H)}_{\le 0}-r\sum_{k=1}^K\Tr\Big(\mathbf{w}_k\mathbf{w}_k^H
\big(\mathbf{Y}_k^*+\frac{\xi_k^*\mathbf{H}_k}{\Gamma_{\mathrm{req}_k}}\big)\Big)+\Delta.
\end{equation}
On the other hand,  since the channel vectors of $\mathbf{g}_m$ and $\mathbf{h}_k$ are assumed to be statistically independent, it follows that by setting $r\rightarrow \infty$, the term $-r\sum_{k=1}^K\Tr\Big(\mathbf{w}_k\mathbf{w}_k^H
\big(\mathbf{Y}_k^*+\frac{\xi_k^*\mathbf{H}_k}{\Gamma_{\mathrm{req}_k}}\big)\Big)\rightarrow -\infty$ and the dual optimal value  becomes unbounded from below. Besides, the optimal value of the primal problem is non-negative for $\Gamma_{\mathrm{req}_k}>0$. Thus,  strong duality does not hold which leads to a contradiction. Therefore, $\mathbf{A}_k^*$ is a positive definite matrix with probability one, i.e., $\Rank(\mathbf{A}_k^*)=N_{\mathrm{T}}L$.

By exploiting (\ref{eqn:lagrangian_gradient}) and a basic  inequality for the rank of matrices, we have
\begin{eqnarray}
\hspace*{-3mm}&&\hspace*{-2mm}\Rank(\mathbf{Y}^*_k)+\Rank\big(\xi_k^*\frac{\mathbf{H}_k}{\Gamma_{\mathrm{req}_k}}\big)\\
\hspace*{-3mm} \notag&\ge &\hspace*{-2mm}\Rank\big(\mathbf{Y}^*_k+\xi_k^*\frac{\mathbf{H}_k}{\Gamma_{\mathrm{req}_k}}\big)=\Rank(\mathbf{A}_k^*)=N_\mathrm{T}L\\
\hspace*{-3mm}&\Rightarrow &\hspace*{-2mm}
 \Rank(\mathbf{Y}^*_k)\ge N_{\mathrm{T}}L-\Rank\big(\xi_k^*\frac{\mathbf{H}_k}{\Gamma_{\mathrm{req}_k}}\big).\notag
\end{eqnarray}
Thus, $\Rank(\mathbf{Y}^*_k)$ is either  $N_\mathrm{T}L-1$ or $N_\mathrm{T}L$. Furthermore, $\mathbf{W}_k^*\ne\mathbf{0}$ is required to satisfy the minimum SINR requirement of  IR $k$ in C1 for $\Gamma_{\mathrm{req}_k}>0$. Hence, $\Rank(\mathbf{Y}^*_k)=N_{\mathrm{T}}L-1$ and $\Rank(\mathbf{W}^*_k)=1$ hold with probability one. In other words,
 the optimal joint beamformer $\mathbf{w}^*_k$ can be obtained by performing eigenvalue decomposition of $\mathbf{W}^*_k$ and selecting the principal eigenvector as the beamformer.

\end{document}